# Diamond as a material for monolithically integrated optical and optomechanical devices

Patrik Rath[1], Sandeep Ummethala[1], Christoph Nebel[2] and Wolfram H.P. Pernice[*,1,3]

[1] Karlsruhe Institute of Technology (KIT), Institute of Nanotechnology (INT), Hermann-von-Helmholtz-Platz 1, 76344 Eggenstein-Leopoldshafen, Germany
[2] Fraunhofer Institute of Applied Solid State Physics (IAF), Tullastr. 72, 79108 Freiburg, Germany
[3] University of Muenster, Institute of Physics, Wilhelm Klemm-Str. 10, 48149 Muenster, Germany

Diamond provides superior optical and mechanical material properties, making it a prime candidate for the realization of integrated optomechanical circuits. Because diamond substrates have matured in size, efficient nanostructuring methods can be used to realize full-scale integrated devices. Here we review optical and mechanical resonators fabricated from polycrystalline as well as single crystalline diamond. We present relevant material properties with respect to implementing optomechanical devices and compare them with other material systems. We give an overview of diamond integrated optomechanical circuits and present the optical readout mechanism and the actuation via optical or electrostatic forces that have been implemented to date. By combining diamond nanophotonic circuits with superconducting nanowires single photons can be efficiently detected on such chips and we outline how future single photon optomechanical circuits can be realized on this platform.

## 1 Introduction

Integrated nanophotonic circuits combine elementary building blocks into complex systems which are interconnected through waveguides. In such photonic integrated circuits light usually takes on the role of information carrier. Yet, another property of light which is often overlooked is momentum, for example, in the form of radiation pressure[1], which can be exploited for optical actuation and light force devices.[2] Harnessing radiation pressure forces allows for implementing a new generation of nanophotonic systems which can be made tunable even if the underlying material platform is passive.[3–5] The successful application of light force tunability critically relies on outstanding optical properties for efficient waveguiding and also outstanding mechanical material properties for realizing high quality resonators. Therefore materials that offer both are particularly attractive.

Diamond has been shown to possess superior optical and mechanical properties and is thus ideally suited for the fabrication of optomechanical devices. Here we present an overview of optical, mechanical and optomechanical components made from diamond. In particular, we consider devices which are made from both single crystal and polycrystalline material. Both material properties and device geometries are discussed for use within the context of integrated optics and optomechanics.

## 2 Optical and mechanical properties of diamond

Diamond provides also excellent thermal and chemical properties. This makes diamond an ideal choice for a wide range of applications where materials are required to operate in aggressive environments exposed to abrasion or chemical attack.[6] Similarly, in situations where extreme levels of power are required to be dissipated, for example in windows for high power lasers, the high thermal conductivity of diamond provides the means to sustain long-term operation. Progress in the synthesis and processing of chemical vapour deposition (CVD) diamond has opened many of these applications by providing polycrystalline diamond material that is available in large areas, in industrial quantities and with the high quality and reproducible properties needed to meet the demanding requirements of the industry.[7]

### 2.1 Optical properties

For applications in integrated optomechanics mostly the optical and mechanical properties of suitable materials are relevant. Efficient waveguiding in photonic devices requires most importantly optical transparency in the spectral region of interest. For waveguides, this necessitates in particular transparency both of the guiding layer, as well as the surrounding optical buffer layers.

* Corresponding author: e-mail wolfram.pernice@uni-muenster.de, Phone: +49 251 83-33633, Fax: +49 251 83-36351

|  | $E_g$ (eV) | TR (μm) | n | E (GPa) | k (W/m K) | ρ (g/cm³) | c (m/s) |
|---|---|---|---|---|---|---|---|
| **Diamond** | 5.47 | 0.22 - 20 | 2.4 | 1100 | 2200 | 3.52 | 17700 |
| **Si₃N₄** | ~5 | 0.3 - 5.5 | 2 | 800 | 33 | 3.24 | 15800 |
| **3C-SiC** | 2.39 | 0.2 - 5 | 2.6 | 390 | 1.4 | 3.21 | 11000 |
| **Sapphire** | 9.9 | 0.17 -5.5 | 1.8 | 340 | 24 | 3.98 | 9200 |
| **AlN** | 6.14 | 0.2 -13.6 | 2.1 | 294 | 150 | 3.26 | 9500 |
| **GaN** | 3.44 | 0.36 - 7 | 2.4 | 294 | 130 | 6.1 | 6900 |
| **TiO2** | 3.5 | 0.42 - 4 | 2.5 | 250 | 10 | 4.26 | 7700 |
| **Si** | 1.12 | 1.1 - 6.5 | 3.5 | 162 | 140 | 2.33 | 8300 |
| **GaP** | 2.26 | 0.54 - 10 | 3.2 | 140 | 100 | 4.13 | 5800 |
| **Ge** | 0.66 | 1.8 - 15 | 4.6 | 132 | 60 | 5.35 | 5000 |
| **GaAs** | 1.42 | 0.9 -17.3 | 3.7 | 116 | 52 | 5.32 | 4700 |
| **ZnO** | 3.4 | 0.37 - | 2 | 110 | 30 | 5.6 | 4400 |
| **SiO₂** | ~9 | 0.38 -2.2 | 1.5 | 95 | 10 | 2.65 | 6000 |
| **InP** | 1.34 | 0.93 - 14 | 3.5 | 89 | 68 | 4.8 | 4300 |

**Table 1 Relevant material properties** for a variety of materials suitable for integrated optomechanics and nanophotonics, namely the band gap $E_g$, transparency range TR, refractive index n, Young's modulus E, thermal conductivity k, density ρ and the sound velocity c.

Wide bandgap semiconductors are particularly attractive for waveguide fabrication (for a comparison see table 1). Besides optical transparency, a high refractive index contrast with respect to the surrounding materials is necessary for nanophotonic components. Because of the high thermal conductivity of diamond[8,9] locally generated heat can be efficiently dissipated across large areas. Together with low absorption due to the absence of two and multi-photon absorption (because of the wide bandgap), this leads to large power handling capacity which can be exploited for diamond nonlinear optics.[10,11] Here diamond is promising for applications in the ultraviolet, visible and infrared spectral region.[10–12] A relatively high nonlinear refractive index ($n_2=1.3\times10^{-19}$ m²/W⁻¹ for visible wavelengths[13]), a large Raman frequency shift (~40 THz) and a large Raman gain (~10 cm/GW at 1-μm wavelength)[10] which is among the highest of available materials for chip-scale nanophotonics, make diamond attractive for on-chip nonlinear optics.

In addition to applications in classical waveguiding devices, high quality diamond layers are attractive because of the possibility to incorporate specific defects into the host material. Color centers in diamond have attracted a lot of attention because they provide promising single photon emission characteristics. Currently known color centers cover a wide range of wavelengths in the visible and near-infrared spectral region.[14] The usefulness of these color centers, such as the nitrogen–vacancy (NV) center, is based primarily on two properties: first, they often feature low electron phonon coupling due to the low density of phonon states i.e. the high Debye temperature. Diamond provides the highest Debye temperature of any material (~2000 K) and is thus an attractive system to explore quantum phenomenon under ambient conditions[15]. Secondly, color centres in diamond are usually found to be long-term stable, even under ambient conditions. This makes them unique among all optically active solid state systems and also provides opportunities for their use as fluorescence marker when embedded in nanoscrystals. In combination with a nanophotonic waveguiding framework color centers could therefore provide the ingredients for waveguide based photonic quantum technologies in a single material platform.

## 2.2 Mechanical properties

With respect to selecting optimal materials for integrated optomechanics, a further constraint is the requirement for outstanding mechanical properties. Diamond offers a very large Young's modulus which directly translates into high mechanical resonance frequencies. These are important for nanomechanical resonators because high frequency operation allows for use of the device under ambient conditions without suffering from significant air damping. This is particularly relevant for the realization of sensing devices. Potential applications include force measurement in the context of scanning probe microscopy[16], mass detection of single molecules[17] and also investigations in fundamental physics[18,19]. Diamond also offers the highest sound velocity among all materials[20] which is important for the realization of high frequency surface acoustic wave (SAW) devices. Furthermore, diamond is a low loss mechanical material. Thermoelastic dissipation is found to be much lower compared to equally sized resonators in materials like silicon. This can be linked to diamond's high thermal conductivity.[21] In comparison with silicon, diamond nanomechanical resonators have been shown to yield substantially lower mechanical dissipation with quality factors (Q factors) in excess of several million at low temperature[22]. Because of the biocompatible properties of diamond[23], such microscale devices also offer possibilities for use within living tissue, with first medical applications in bionic-eye research already under way[24].

## 2.3 Polycrystalline versus single crystalline diamond

When implementing photonic circuits in diamond one can either work with single crystalline diamond (SCD) or polycrystalline diamond (PCD), which differ in terms of material properties. While some of the advantageous properties of diamond as listed in table 1 are largely preserved when diamond is not single- but polycrystalline (namely the refractive index, the Young's modulus, the density and the sound velocity), others like the transparency range and the thermal conductivity are reduced, mainly due to grain boundaries which incorporate sp2-carbon and larger amounts of dopants than bulk diamond.[8,14] Thin PCD layers grown to thicknesses between 0.1 and 10 μm are generally inferior in terms of their material properties and background impurity levels. On the other hand, PCD grown to thicknesses greater than 100 μm can have material properties quite comparable to SCD because the grain size becomes much larger in this case. Nevertheless, the thermal conductivity even of a thin film of PCD is still larger than the thermal conductivity of any other material listed in table 1[25,26]. The incorporation of isolated color centers in PCD is currently hindered by background fluorescence which is caused by grain boundaries. Especially strain and impurities can deteriorate the properties of color centers[27]. Improved growth procedures might enable the combination of color centers and photonic components from PCD in the future. When it comes to material availability and maturity of fabrication procedures there is a striking difference between PCD and SCD: because SCD can only be grown on SCD substrate, no wafers with SCD thin films are currently available. A range of different transfer and undercut methods, which we describe in detail in section 4.1, are being explored in order to create photonic components from SCD. PCD on the other hand can be grown on different substrates as thin films via chemical vapor deposition (CVD). This leads to wafer-scale availability and enables the use of fabrication processes employed for other semiconducting thin films, as illustrated in Fig. 7. Therefore up-to-date more advanced device geometries in integrated optomechanics have been being implemented on PCD, as will be discussed in the following sections.

### 3 Integrated Optomechanics

When used within the context of optomechanics, freestanding diamond devices are actuated via on-chip radiation pressure forces[28]. The resulting forces have proven to be effective in manipulating nanoscale objects when moving to dimensions on the order of the optical wavelength and below. This results from the close match between the crosssection of a waveguide and a nanomechanical resonator which leads to efficient overlap between optical and mechanical modes.

Nano-optomechanical devices have been used in particular for detecting the motion of objects embedded within a resonant cavity[2,29,30]. In its simplest form such a cavity-optomechanical system consists of an electromagnetic resonator with its resonance frequency dispersively coupled to the position of a movable mirror. In such a cavity-based scheme, a narrowband optical source is used to drive the system. The mechanical motion of the resonator resulting from radiation pressure then translates into modulation of the stored intra-cavity electromagnetic field. At the same time, the filtering properties of the cavity imprint the mechanical motion onto the electromagnetic signal which can be detected in the spectrum of the cavity output.

Various cavity designs can be exploited for this purpose, with optical Q factors exceeding $10^6$.[11] The resonant enhancement of the pump's radiation pressure further leads to strong back-action effects on the mechanical resonator. These modify the dynamic mechanical and optical properties of the coupled system. Dynamical back-action effects can lead to optical stiffening of the mechanical structure[1,31–33], damping or amplification of the mechanical motion[1,34–36], or electromagnetically induced transparency[37–39]. Cavity-optomechanical systems demonstrating near quantum-limited position read-out and strong radiation pressure back-action have been realized both in the optical[40,41] and the microwave frequency domains[42,43].

In integrated nanophotonic circuits out-of-plane radiation pressure forces are not ideally suited for actuation because of the planar layout of the photonic components. Instead gradient optical forces can be used which result from modifications of the propagating modal energy due to the presence of a movable mechanical structure[4,3]. This directly translates into a change in the effective refractive index and thus a measurable phase shift induced by nanomechanical motion. The introduction of refractive index changes is particularly useful because it can be readout with phase-sensitive detection schemes.[44] Several possible options are available in integrated optical circuits, including cavities and interferometers, such as the Mach-Zehnder Interferometers described in section 6.2. When the properties of the waveguiding material are known, such elements can be readily implemented on different material platforms, including diamond.

### 4 Diamond integrated optical circuits

In order to develop large scale diamond photonic circuits the individual nanophotonic devices which are connected together by nanophotonic waveguides need to be developed and optimized. The resulting toolbox, from which integrated optical circuits can be assembled, can be classified into the following categories:

**(1) Waveguides** are the most simple, but also most important building blocks. They consist of stripes of dielectric material which due to a refractive index contrast to the surrounding media can confine light and host optical modes, in analogy to optical fibers. Waveguides are used to route light across the photonic chip and are furthermore used to implement beam splitters in the form of directional couplers or Y-splitters, as shown Fig. 1(i).

**(2) Fiber-to-chip couplers**, which allow to transfer light into and out the waveguides and connect them to off-chip components such as light sources or detectors are either implemented in plane by butt coupling to the facet of a polished waveguide or realized via grating structures, which enable out of plane access to waveguides. Such grating couplers as shown in Fig. 1(ii) give more design flexibility for photonic circuits and can increase the integration density due to a small device footprint.

**(3) Optical resonators or cavities**, such as disk resonators, ring resonators (Fig. 1(iii)), one-dimensional (1D) and two-dimensional (2D) photonic crystals cavities (PhC) which are characterized by their optical Q factor.

**(4) On-chip interferometers**, such as Mach-Zehnder Interferometers (MZI), as shown in Fig. 1(iv), which are useful for translating phase changes into intensity changes and find applications for chip-based sensors and modulators.

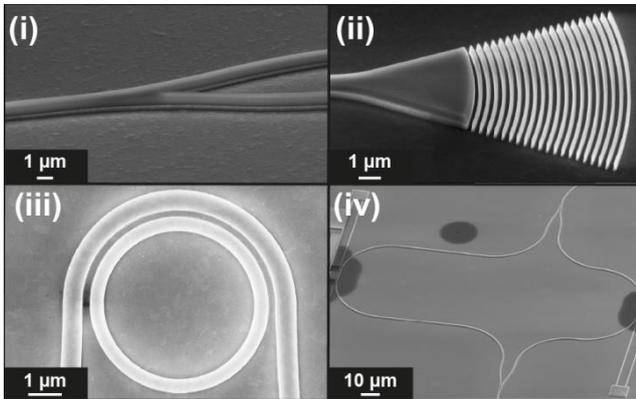

**Figure 1 Examples of essential components in diamond integrated optical circuits:** (i) Waveguide and beam splitter (ii) Grating coupler (iii) Ring resonator and (iv) Mach-Zehnder Interferometer, fabricated from PCD thin film.

Because there is up to date no method for directly growing SCD thin films on buffer layers with lower refractive index such as silica, diamond as a substrate for integrated optics has been largely unexplored until recent years. Demonstrations of the components outlined above so far have been mainly focused on single components from SCD and PCD opposed to larger photonic circuits. A particular effort has been dedicated to developing optical cavities with high Q factors. An overview of literature on diamond optical resonators at room temperature is given in Table 2. These data sets have almost exclusively been published within the last 5 years, underlining the fast pace at which this field of research is currently advancing. Table 2 is sorted into two categories: demonstrations of cavities for visible wavelengths (1-17) and demonstrations for infrared wavelengths (18-23). Visible wavelength cavities are of interest for enhancing the emission from color centers, while near-IR cavities primarily are developed for applications in the telecommunication regime. For both wavelength regimes the resonators are first sorted by their type and then sorted by their optical quality factor Q. For every resonator type the demonstration which up to date shows the highest quality factor is marked in blue.

|   | Resonator type | Diamond type | Wavelength (nm) | Quality Factor Q | Fabrication strategy | Ref |
|---|---|---|---|---|---|---|
| 1 | 1D PhC | SCD | 700 | 9900 | Thin down + EBL (1) | 45 |
| 2 | 1D PhC | SCD | 671 | 6000 | Thin down + EBL (1) | 46 |
| 3 | 1D PhC | SCD | 648 | 5100 | EBL + Angle Etching (3) | 47 |
| 4 | 1D PhC | SCD | 762 | 3060 | EBL + Angle Etching (3) | 48 |
| 5 | 1D PhC | SCD | 630 | 700 | Focused ion beam (5) | 49 |
| 6 | 1D PhC | SCD | 639 | 220 | Focused ion beam (5) | 50 |
| 7 | 2D PhC | SCD | 670 | 1200 | Focused ion beam (5) | 51 |
| 8 | 2D PhC | SCD | 778 | 1000 | Liftoff + EBL (1) | 52 |
| 9 | 2D PhC | PCD | 590 | 585 | Deposition + EBL | 53 |
| 10 | 2D PhC | SCD | 810 | 450 | Focused ion beam (5) | 49 |
| 11 | 2D PhC | SCD | 740 | 320 | Focused ion beam (5) | 54 |
| 12 | Disk | SCD | 637 | 3000 | Liftoff + EBL (2) | 55 |
| 13 | Disk | SCD | 737 | 2200 | Liftoff + EBL (2) | 56 |
| 14 | Disk | PCD | 637 | 100 | Deposition + EBL | 57 |
| 15 | Ring | SCD | 637 | 59000 | EBL + Angle Etching (3) | 47 |
| 16 | Ring | SCD | 686 | 12600 | Thin down + EBL (1) | 58 |
| 17 | Ring | SCD | 638 | 12000 | Thin down + EBL (1) | 59 |
| 18 | 1D PhC | SCD | 1609 | 183000 | EBL + Angle Etching (3) | 47 |
| 19 | 2D PhC | PCD | 1622 | 6500 | Deposition + EBL | 60 |
| 20 | Disk | SCD | 1552 | 115000 | EBL + undercut etching (4) | 61 |
| 21 | Ring | SCD | 1545 | 1000000 | Thin down + EBL (1) | 11 |
| 22 | Ring | SCD | 1650 | 151000 | EBL + Angle Etching (3) | 47 |
| 23 | Ring | PCD | 1550 | 11000 | Deposition + EBL | 62 |

**Table 2 Overview of diamond optical resonators measured at room temperature.** Different resonators are sorted by their resonance wavelength and resonator geometry. For each resonator type at a specific wavelength range the device with the highest Q factor is highlighted in blue.

**4.1 Single crystalline photonic components**
Fabricating photonic circuits in SCD is challenging due to the fact that SCD can only be grown homo-epitaxial on existing SCD. Therefore the size of available SCD templates is limited to some 10s of $mm^2$. In order to realize a thin film or a freestanding diamond membrane, as necessary for photonic circuit manufacture, different methods have been explored, which can be categorized as "thin-down" strategy, "lift-off" technique, "angle-etching" procedure, as well as "undercut" etching and focused ion beam milling. Table 2 lists which of the devices have been fabricated using one of the following methods:

**(1) The "thin-down" strategy** starts with a diamond piece with a thickness of tens of microns, often acquired from commercial suppliers, which is then thinned down by dry etching to a thickness of some hundreds of nanometres. Thickness variations of the thick bulk material are directly transferred into the thin film, so one ends up with a wedge-like layer, unless further precautions are taken.[63] Next, electron beam lithography (EBL) and dry etching are used to create the photonic structures via lithographic structuring.

**(2) The "lift-off" technique** circumvents the challenge of the inhomogeneous layer thickness by irradiating the diamond with ions.[64,65] During an annealing step the damaged layer graphitizes and can be selectively removed. After regrowing a pristine diamond layer, removing any damaged diamond and transferring it to a different substrate, structures can be patterned by EBL and dry etching.[55]

**(3) The "angle-etching" procedure** consists of first patterning resist by EBL and then transferring the resist structures into the bulk diamond by angled anisotropic plasma etching.[66] This method allows to directly structure freestanding devices into bulk diamond without the need for previous thin-down, but restricts the design freedom for photonic components as for a given device width the height of the resulting structure is automatically fixed by the etching angle.

**(4) The "undercut" etching method** consists of patterning resist by EBL and then transferring the resist structures into the bulk diamond by a series of steps of EBL, coating with $Si_3N_4$ and different dry etching plasmas, including a quasi-isotropic reactive ion etching step.[61,67] This newly emerging method allows to fabricate vertical side walls as opposed to "angle-etching", while still undercutting the structures, in order to get freestanding devices on top of bulk diamond.

**(5) Focused ion beam (FIB) milling** uses typically $Ga^+$-ions in order to directly cut into a given material. This method allows the direct patterning of diamond without the need for generating a mask[49–51,68,69], but has drawbacks concerning damage, implanted ions and scalability of the process. This might in the future be circumvented by electron-beam induced etching (EBIE)[70,71] which has been shown to have the potential to directly structure diamond without the need of an ion beam.

Figure 2 shows an overview of SCD optical cavities. It is interesting to note that the best performing device in each resonator category (best Q in the visible: 1D PhC: 9900, 2D PhC: 1200, disk: 3000, ring: 59000) have to date been achieved each with a different fabrication method. Methods 3-5 cut structures directly into bulk diamond, which raises the question of scalability of such integrated optical circuits. Methods 1-2 lead to thin membranes which, besides having a fairly limited size, can be transferred onto a different substrate, such as oxidized silicon wafers. Afterwards, the buffered optical devices can be processed similar to other materials for integrated optics which as opposed to SCD are already today available as homogeneous thin films of large area.

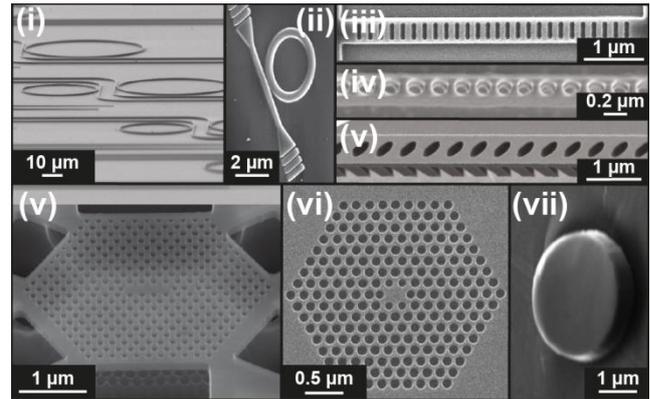

**Figure 2 Examples of SCD optical cavities:** (i,ii) Ring resonators[11,59], (ii,iv) 1D PhC cavities[45,47,72], (v,vi) 2D PhC cavities[51,73], and (vii) a disk resonator[56].

### 4.2 Polycrystalline photonic components

Differing from SCD polycrystalline diamond can be readily deposited as homogeneous thin films via plasma enhanced CVD[74]. Slurry based chemo-mechanical polishing allows to reduce the intrinsic roughness of the as-deposited diamond thin film significantly.[75] Therefore the fabrication methods for this type of material are scalable because only lithography, dry etching and wet etching are needed and no transfer of membranes or exotic procedures for undercutting diamond components into bulk diamond are required. Opposed to SCD fabrication procedures, this allows to readily structure full photonic networks from PCD, incorporating all of the components categorized in Section 4. Different types of optical cavities[53,57,60,76], waveguides of several millimetre length[62] and integrated Mach-Zehnder Interferometers[44,77] have been demonstrated to date and Fig. 2 shows examples of such components.

While PCD currently has disadvantages for applications in the visible wavelength regime due to grain boundaries and higher level of dopants, there is ongoing research concerning the incorporation of color centers into PCD[78,79]. Further advances in material deposition and characterization might pave the way for applications of PCD thin films also at visible wavelengths. Already the performance of 2D PhC cavities, with Q factors up to 6500 at 1550 nm[60] and ring resonator, with Q factors up to 11.000 at 1550 nm[62] show the potential of PCD diamond layers for integrated optics in the near infrared spectral region, especially when interfacing it with nanomechanical components.

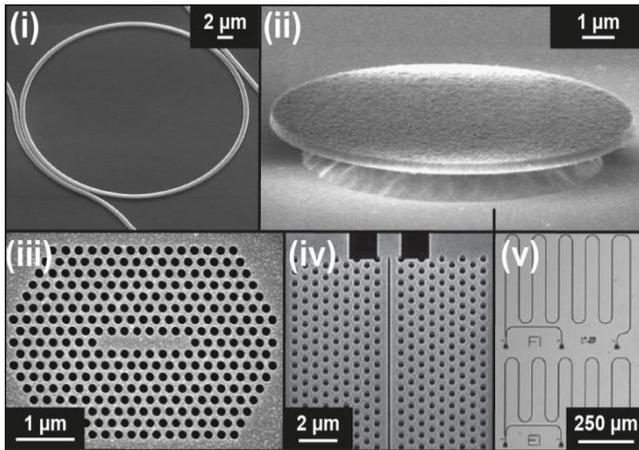

**Figure 3 Examples of PCD optical components:** (i) Ring resonators[76], (ii) Disk resonator[57], (iii, iv) Photonic Crystals[53,60], (v) Millimeter long waveguides[62].

### 5 Diamond nanomechanical resonators

For use as optomechanical transducers and sensing devices nanomechanical resonators are particularly interesting because of their very small mass and thus high responsivity to changes in their local environment. For use within nanophotonic circuits, the mechanical resonators need to be compliant with a waveguide architecture. For this purpose some of the most common nanomechanical resonator geometries are cantilevers, which are clamped only on one end, or nanoscale beams which are clamped on both ends. Such structures can be naturally realized by partially releasing on-chip waveguides through removing the underlying substrate[44,4,80,81]. These structures can also be embedded in photonic cavities such as ring and race-track resonators. Nevertheless, the transition from unreleased waveguide sections to free-standing structures induces optical losses and therefore reduces the overall optical quality factor of the resonator.

To circumvent this issue, centrosymmetric mechanical resonators, such as wheel[82,83] and disk resonators[84,85] can be employed, which have the advantage that they can be easily designed to host mechanical resonances and optical resonances simultaneously. In this case the entire optical resonator is made free-standing, thus avoiding degradation of the optical resonator performance. Embedded optomechanical resonators have been realized in a variety of material systems including silicon, silicon nitride, aluminium nitride[86] and silicon carbide[87]. In particular silicon carbide and aluminium nitride optomechanical resonators have allowed for pushing the mechanical response to the ultra-high frequency and super high frequency regime[87,88]. For increasing mechanical resonance frequencies, the high Young's modulus of diamond is very attractive, as can be seen by comparing it to other materials in table 1. The large Young's modulus allows for pushing the resonance frequencies by a factor of three compared to single crystal silicon resonators of the same geometry.

| | Resonator type | Diamond type | V ($\mu m^3$) | f (MHz) | Quality Factor Q | Q*f (GHz) | T(K) | Ref |
|---|---|---|---|---|---|---|---|---|
| 1 | Dome resonator | SCD | 4.2 | 27 | 4000 | 108 | 300 | 89 |
| 2 | Doubly clamped beam | PCD | 5.3 | 3.8 | 11200 | 42.6 | 300 | 44 |
| 3 | Cantilever | SCD | 6.8 | 4.6 | 50300 | 229.9 | 300 | 90 |
| 4 | Doubly clamped beam | SCD | 11 | 6.7 | 19000 | 127.3 | 300 | 90 |
| 5 | Doubly clamped beam | SCD | 14 | 3.1 | 199000 | 616.9 | 300 | 67 |
| 6 | Dome resonator | SCD | 15 | 50 | 2000 | 100 | 300 | 91 |
| 7 | H-resonator | PCD | 30 | 120 | 1300 | 149.5 | 300 | 92 |
| 8 | H-resonator | PCD | 34 | 6.5 | 28800 | 185.8 | 300 | 77 |
| 9 | Cantilever | PCD | 600 | 0.32 | 116000 | 36.9 | 300 | 93 |
| 10 | Cantilever | SCD | 1530 | 1.4 | 54000 | 75.6 | 300 | 94 |
| 11 | Cantilever | SCD | 1900 | 0.032 | 412000 | 13.2 | 300 | 22 |
| 12 | Cantilever | SCD | 2400 | 0.73 | 338000 | 246.7 | 300 | 94 |
| 13 | Cantilever | PCD | 3590 | 0.046 | 86000 | 4 | 300 | 21 |
| 14 | Doubly clamped beam | PCD | 3.6 | 9.39 | 40000 | 375.6 | 3 | 95 |
| 15 | Fishbone | PCD | 6.24 | 1440 | 8660 | 12470 | 1.1 | 96 |
| 16 | Fishbone | PCD | 12.8 | 631 | 23200 | 14639 | 1.1 | 96 |
| 17 | Doubly clamped beam | SCD | 13.5 | 3.1 | 720000 | 2232 | 5 | 67 |
| 18 | Cantilever | SCD | 1900 | 0.032 | 1510000 | 48.3 | 3 | 22 |

**Table 3 Overview of diamond nanomechanical resonators.** Different resonators sorted by their volume, measured at room temperature (1-13) and at Cryogenic temperatures (14-18).

Table 3 shows an overview of SCD and PCD mechanical resonators, sorted by the resonator's volume V, both for room temperature measurement and for cryogenic conditions. All measurements were taken in vacuum at pressures below $10^{-5}$ Torr. From published data the resonator with either the highest frequency or the highest Q factor was selected. As in the case of silicon devices, diamond resonators in the GHz frequency regime have been demonstrated[96] and higher frequency operation can be expected in the near future. In the following we compare the performance of single and polycrystalline diamond resonators.

### 5.1 Single crystalline nanomechanical components

Single crystal materials are usually expected to yield superior mechanical performance due to a perfect crystal structure and low defect density which can lead to dissipation. A variety of diamond single crystal mechanical resonators is shown in Fig.4. Ultrahigh mechanical Q factors have been demonstrated with cantilever devices[22], exceeding one million at room temperature for a cantilever thickness in the micrometer range. Even higher Q factors approaching 6 million become accessible when moving to cryogenic temperatures. Interestingly, the demonstrated Q factors are about 2 orders of magnitude higher than in polycrystalline structures of the same dimensions.

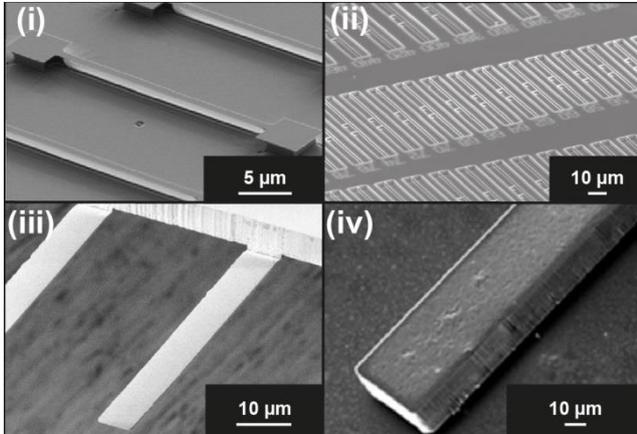

**Figure 4 Examples of SCD nanomechanical devices:** (i,ii) Doubly clamped beams[67,90] and (iii,iv) Singly clamped cantilevers[22,94].

A particular interest in high Q SCD cantilevers occurs because they can be coupled to embedded color centers[97–99] for example for local strain sensing. This has recently become feasible by coupling a mechanical resonator to an embedded single spin through lattice strain. Such an approach will allow for using NV centers embedded in single-crystal diamond nanomechanical resonators for hybrid systems in the quantum regime. Potential examples include spin-induced oscillator sideband cooling, spin squeezing, or ultrafast, mechanical spin driving. The first demonstrations of such applications will be discussed in section 5.3.

While SCD cantilevers provide very high mechanical performance, their out-of-plane motion is not ideally suited for operation within integrated optical circuits. Instead freestanding nanostrings in the form of doubly-clamped beams are preferred which couple evanescently to either waveguides or other readout approaches. Doubly clamped beams lead to slightly lower Q factors, with demonstrated values of about 200.000 at a resonance frequency of 3 MHz.[67] As explained in section 4.1 the angle-etching and the undercut method allow for creating free-standing resonators from bulk diamond substrates without having to use buffer layers or heterostructures. These approaches hold promise for realizing high quality mechanical structures from optimized diamond template material, but their integration into photonic circuits would be challenging as all photonic components would need to be isolated from the underlying diamond bulk material and therefore the full photonic circuitry would need to be underetched and freestanding.

### 5.2 Polycrystalline nanomechanical components

Despite impressive progress in the preparation of SCD diamond devices, the limited substrate size will impose stringent limitations, in particular with respect to real world applications. In this respect PCD films that can be realized on a waferscale provide a promising alternative. PCD has been employed for realizing Micro-Electro-Mechanical Systems (MEMS)[100–102] due to its large sound velocity (>18000 m/s), the small friction and wear and the large Young's modulus, which has been shown to be in par with the one of SCD[103,104].

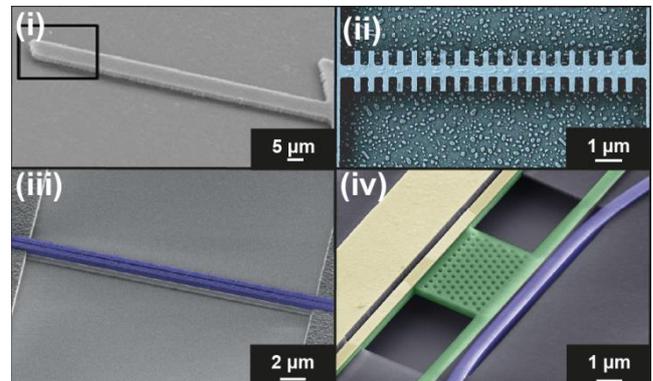

**Figure 5 Examples of PCD nanomechanical devices:** (i) Singly clamped cantilevers[93] (iii) Doubly clamped beams[44] or more exotic structures, such as (ii) the fishbone resonator[96] and (iv) the H-resonator[92].

Polycrystalline diamond resonators have been experimentally shown to reach Q*f product surpassing $10^{13}$ Hz[96] which is currently a record value for any diamond nanomechanical resonator, as compared in table 3. The advantage of the outstanding mechanical properties can furthermore be combined with other attractive material properties, for example the fact that diamond can be piezoresistive when doped p-type[105] or that superconducting nano-mechanical diamond resonators are possible.[95] Furthermore, it has been shown that PCD nanomechanical resonators can be monolithically combined with integrated optical circuits and operated via optical forces[44] or electrostatic forces on chip[92]. This way efficient all-optical driving can be achieved under high vacuum conditions, or strong electrostatic driving for use under ambient conditions with direct applications in chip-based sensing. A selection of nanomechanical devices prepared from PCD diamond is shown Fig.5.

Even though the currently demonstrated mechanical Q factors are somewhat smaller than for SCD devices of the same geometry, routes for further increasing the Q factors are available. One avenue consists in strain engineering by inducing tensile material stress into the diamond layer during growth or through geometric strain concentration. In addition, improvements in the synthesis of suitable diamond thin films might lead to reduced mechanical dissipation in fabricated optomechanical devices.

### 5.3 Applications of diamond nanomechanical sensors

Nanomechanical devices made from diamond are progressing towards a range of applications. To date, a particular interest lies in the availability of the nitrogen vacancy defect center within movable mechanical structures. A single NV spin embedded in a single crystal diamond cantilever can be accessed optically via a confocal microscope for initialization and readout as illustrated in Fig. 6a.[98] Microwaves are used for pulsed spin manipulation. As pointed out above, this configuration allows for local strain sensing through strain-mediated coupling of a single diamond spin to a mechanical resonator. Through quantum control of the spin, the axial and transverse strain sensitivities of the nitrogen–vacancy ground-state spin have been determined with a strain sensitivity of $3 \times 10^{-6}$ strain $Hz^{-1/2}$. This allows in particular for measuring a motional change of 7 nm within 1 second integration time. Corresponding experiments were also reported in[97].

The NV defect centre in diamond also provides potential applications in nanoscale electric and magnetic-field sensing. This can be achieved by positioning a single nitrogen-vacancy centre at the end of a high-purity diamond nanopillar, which is used as the tip of an atomic force microscope[106]. By scanning a single NV centre within nanometres of a sample surface magnetic domains with widths of 25 nm can be discriminated. The magnetic field sensitivity of 56 nT $Hz^{-1/2}$ at a frequency of 33 kHz provides promising possibilities for magnetometry operation even under harsh environmental conditions.

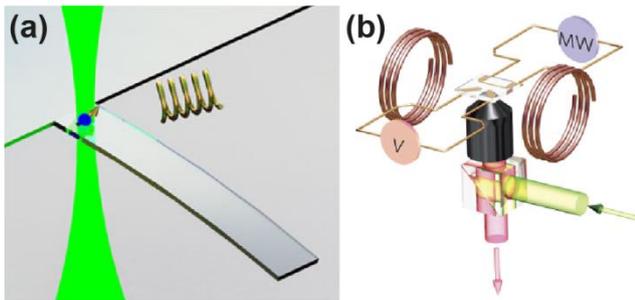

**Figure 6 Sensing Concepts:** (a) Dynamic strain-mediated coupling of a single diamond spin to a mechanical resonator.[98] (b) Electric-field sensing using single diamond spins.[107]

The NV center further allows for three-dimensional electric-field measurements[107]. As illustrated in Fig. 6b a confocal set-up in combination with Helmholtz coils allows for magnetic field alignment, while a microstructure on the diamond sample is used to create the electric field and to couple in microwaves. In this configuration a photon shot-noise limited a.c. electric-field sensitivity of $202 \pm 6$ V $cm^{-1}$ $Hz^{-1/2}$ was achieved, allowing the detection of the electrostatic field produced by a single elementary charge located at a distance of 150 nm from the sensing NV spin within one second of averaging. These way individual charges can be discriminated with nanometre spatial resolution under ambient conditions. Using NV centers for sensing applications also allows for switching between electric- or magnetic-field detection modes as a universal detector system. Potential integration with nanomechanical structures would then enable versatile scanning probe measurements for a wide range of fields.

Even though these sensing concepts show many advantages, so far they have not been interfaced with integrated optical circuits, partially due to the challenging fabrication described in section 4. A main obstacle for this research direction remains to find a way to bring together both superior material quality and scalable fabrication.

## 6 Diamond integrated optomechanics

While there has been much progress in recent years concerning the fabrication of mechanical and optical devices of high quality out of SCD, there has been up to our knowledge no example of a full optomechanical system in SCD. Using PCD on the other hand there have been demonstrations of that kind[44,92], which show the potential of diamond as a material for integrated optomechanics. The following section aims to summarize the state of research in this area.

### 6.1 Fabrication of diamond optomechanical circuits

A general process flow for the fabrication of diamond optomechanical circuits from polycrystalline diamond thin films is depicted in Fig. 7. For such optomechanical applications diamond-on-insulator wafers (DOI) are used which typically consist of a thin polished[75] PCD layer of 600 nm or less, grown via plasma enhanced CVD, on top of 2000 nm of oxidized silicon on a standard silicon wafer, as illustrated in Fig. 7(1).

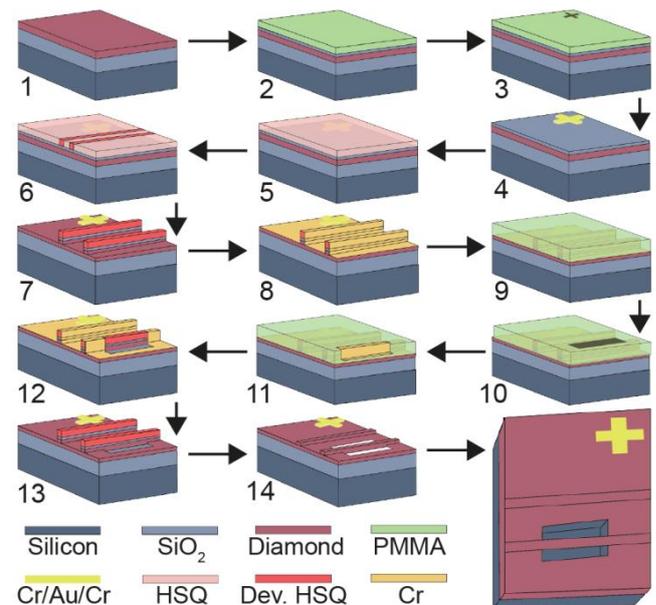

**Figure 7 Process flow for fabrication** of optomechanical circuits with photonic and mechanical resonators on a diamond-on-insulator wafers: First metal electrodes and alignment markers are realized through a lift-off technique, using the positive tone resist PMMA (1- 4). Next the photonic circuits and the mechanical res-

onators are defined in negative tone resist HSQ (5-6) and transferred into diamond via a partial dry etching step (7). Then a thin layer of chromium is deposited onto the sample (8), which acts as a hard mask in the following steps. Finally opening windows for the future freestanding components are written into PMMA (9-11) and transferred into the chromium layer via wet etching (12). The chromium hard mask then allows to fully etch the diamond layer within the opening windows (13) revealing the underlying silicon oxide layer. The mechanical resonators are then released (14) using diluted hydrofluoric acid.

The integrated photonic circuit fabrication is done by planar fabrication employing multiple steps of EBL, wet and dry etching, as illustrated in Fig. 7 and described in detail elsewhere.[77,92] For research purposes, this fabrication procedure is typically employed on wafer dies of 15x15 mm size, but for commercial applications it could be readily employed on DOI wafers, with diameters currently possible up to 6 inch. A false color scanning electron microscope (SEM) image of an example of a final optomechanical circuit is shown in Fig. 8a. All developed fabrication steps could be directly employed on DOI wafers with a single crystalline instead of polycrystalline diamond layer once these become available.

### 6.2 On-Chip read-out of mechanical motion

Typical photonic circuits, as described in section 4.2 employ focusing grating couplers[108,109] in order to couple light from aligned external optical fibres into and out of the photonic circuits. On chip interferometers, such as Mach-Zehnder interferometers (MZI), shown in Fig. 8a, are useful building blocks in integrated optics in order to allow phase-sensitive measurements. The normalized transmission spectrum of an integrated MZI is shown in Fig. 8b showing the expected interference fringes and the envelope given by the grating couplers.

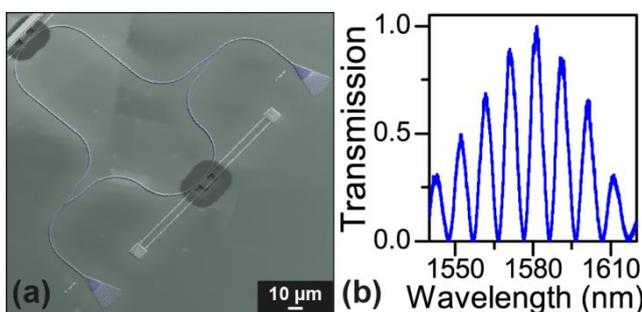

**Figure 8 Mach-Zehnder interferometers:** (a) Scanning electron micrograph of an on-chip polycrystalline diamond MZI. b) Normalized transmission of the MZI showing the expected interference fringes and an envelope given by the spectral profile of the grating couplers.

In the context of integrated optomechanics, MZIs enable measuring mechanical displacement with high sensitivity. The working principle is to interface mechanical components with the waveguides of the MZI, such that the effective refractive index $n_{eff}$ of the optical mode guided in the waveguide depends on the displacement of the mechanical resonator. Creating this dependence is possible by placing a mechanical resonator in the evanescent field of a waveguide, as for example in the case of a so called H-resonator[92,110], shown in Fig. 5(iv), or by using a so-called slot waveguide geometry[44,111], as shown in Fig. 5(iii) where the mode is guided mainly in the air between two dielectric beams, which themselves act as mechanical oscillators.

For both geometries a displacement of the freestanding mechanical structure, often expressed as a change in the separation $g$ between between the waveguide and a second counterpart, leads to a change in $n_{eff}$, which corresponds to a change in the speed of light along the waveguide. This change leads to a phase shift in one interferometer arm compared to a second interferometer arm and hence changes the interference condition at the final beam splitter. As a result a mechanical displacement leads to a change in the intensity of the light transmitted at the output of the MZI, which can be recorded with a fast photodetector.

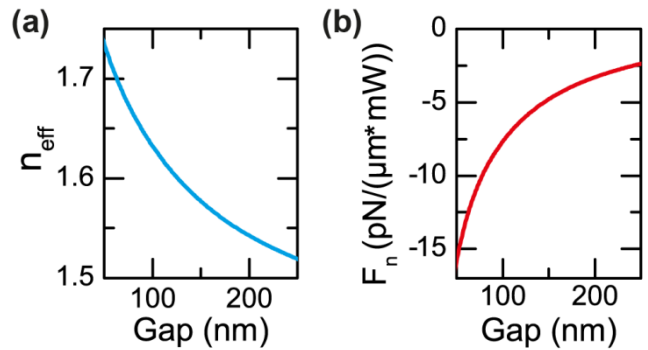

**Figure 9 Relation between mechanical displacement and the effective refractive index and the resulting optical gradient force:** (a) The calculated effective refractive index $n_{eff}$ for a diamond slot waveguide with two dielectric beams of 400 nm width and 600 nm height, depending on the size of the gap between the dielectric beams. (b) The optical force $F_n$, resulting from the gradient of the refractive index, normalized to the beam length L and optical power P.

Fig. 9a shows the results of a finite element method (FEM) simulation of the effective refractive index $n_{eff}$ for a diamond slot-waveguide, consisting of two dielectric beams of 600 nm height and 400 nm width. When one of the beams is displaced, the gap size g changes, leading to a rapid decrease in $n_{eff}$, which translates via the MZI into an intensity change as explained above. This rapid change in $n_{eff}$ enables detecting the thermomechanical motion of the mechanical beams, which allows to quantify the displacement sensitivity, amounting to $11 \times 10^{-15} m/\sqrt{Hz}$ using 2 mW of optical power inside the waveguide.[44] This is a useful feature of the optomechanical system, as it enables to extract important system characteristics such as the mechanical quality factor and the displacement sensitivity, by exploiting the thermal energy of the system, without the need for an active actuation of the mechanical oscillator.

### 6.3 Driving mechanical motion by optical forces

Besides enabling the detection of the mechanical motion, the change of the effective refractive index with a mechanical displacement as described in section 6.2 also implies an attractive gradient optical force occurring in such an optomechanical device. In the case of the slot waveguides a normalized optical force between the two dielectric beams occurs, which amounts to

$$F_n = \frac{F_{opt}}{PL} = \frac{\frac{n_g}{cn_{eff}}}{\frac{\partial n_{eff}}{\partial g}}$$

where $n_g$ is the group refractive index, $n_{eff}$ the effective refractive index, $c$ the vacuum speed of light and the partial derivative is carried out with respect to the separation $g$ between the beams. The force is normalized to the optical power $P$ and the beam length $L$, and thus given in units of $\frac{pN}{\mu m\, mW}$. Note that this means that the applied optical force is proportional to the optical power inserted into the waveguide.

The described optical force allows for manipulating the mechanical motion of an integrated mechanical resonator purely by the force of light which can be demonstrated using a pump-probe measurement scheme[44]: the intensity of light of a chosen wavelength from one laser is modulated at a given frequency in the MHz-GHz range (pump). This modulated intensity leads to a modulated optical force, which drives the mechanical motion of the resonator. A weak continuous wave laser of a different wavelength then the pump light is used to detect the motion (probe). Spectral filtering makes sure that only the modulated probe light is being detected, while the pump light is orders of magnitudes weaker such that it can be neglected.

### 6.4 Optically decoupled nanomechanical resonators

Employing optical forces in the same waveguide to coherently drive mechanical motion is an elegant concept, but it requires the use of optical filters with high extinction ratio in order to separate pump and probe light and avoid cross talk. A way to avoid this is to separate the mechanism of the driving force from the read out signal, for example by using a 2D photonic crystal (PhC). Fig. 5(iv), shows the implementation of such a PhC mirror, which optically isolates the two sides of a mechanical resonator.

The employed geometry, called H-resonator[77,110] consists of two nanobeams joined together through a central plate. This geometry has many degrees of freedom in its design which, opposed to the slot waveguide geometry discussed earlier, allows to conveniently tune the resonance frequency. Furthermore, the H-resonator provides improved mechanical stiffness and is therefore also suitable for the realization of long resonator structures.

In general, the H-resonator supports several optical modes. When brought into the vicinity of an optical readout waveguide, as shown in Fig. 10a, light couples evanescently from the waveguide into the H-resonator. In this geometry light injected into the readout waveguide would be coupled into the central part of the H-resonator and strongly attenuated which is undesirable. Therefore photonic crystal mirrors are employed (Fig. 10b) to optically separate the upper and lower mechanical arms and avoid the attenuation of the readout light.

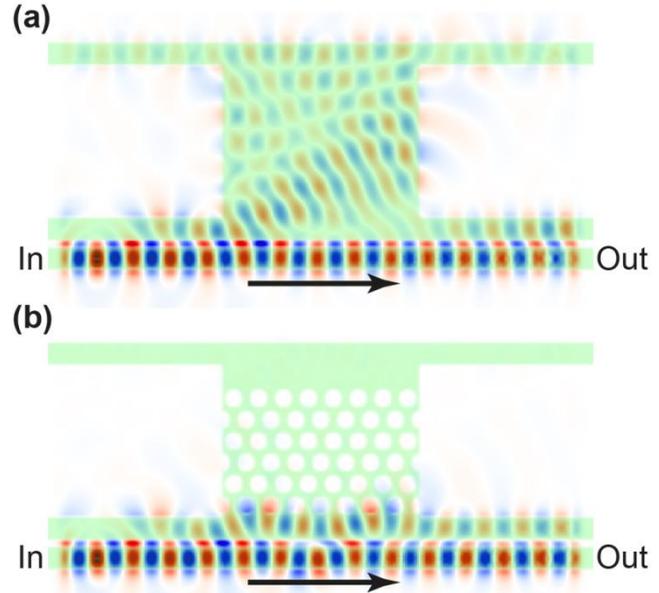

**Figure 10** Simulated mode pattern for an H-resonator which is evanescently coupled to a waveguide at the bottom (a) without and (b) with a photonic crystal mirror inside the H-resonator. The PhC mirror optically decouples the bottom and top arms

The readout waveguide is designed as a single mode waveguide for TE-like polarisation. Using 3D finite-difference time-domain (FDTD) methods the leakage of light into the H-resonator can be minimized. Simulated field patterns, illustrated in Fig. 10, show how the PhC mirror restricts light propagation to one side of the mechanical resonator. The inset of Fig. 11 shows the geometry of the PhC lattice. For a 600 nm thick freestanding diamond membrane the simulations yield an optimal photonic crystal lattice constant of 600 nm with a fill-factor of the hexagonal lattice of 70 %. As shown in Fig. 11 a photonic crystal slab with this geometry shows a bandgap for TE-like polarization that spans a normalized frequency range which corresponds to wavelengths of 1400-1770 nm. This covers the important telecom C-band, which is the target wavelength range for the employed photonic circuitry for optomechanical motion read-out.

### 6.5 Driving mechanical motion via on-chip electrodes

The PhC mirror of the H-resonator design allows for optically separating the two arms of the mechanical resonator, such that one side can be used for implementing a mechanism of driving the mechanical motion, while the other side is interfaced with the optical readout mechanism. For this purpose an electrostatic drive as in MEMS devices can now be combined with the advantage of high sensitivity of the optical readout scheme. This creates a new electro-optomechanical system, without inducing optical losses due to the added metal components. The device consists of a metal electrode directly patterned on the top of the arm of the freestanding diamond mechanical resonator which is on the opposite side of the resonator as the readout-waveguide, as shown in Fig. 5(iv). The movement of the resonator arms is translated into an intensity modulation at the output of a MZI, as explained in section 6.2.

Figure 8a shows the full optomechanical circuit, including the grating couplers, the MZI and the H-resonator. The resonator can be driven by the combination of a DC voltage and a radio-frequency (RF) signal which is applied between the electrode on the freestanding mechanical resonator and a counter electrode on the fixed substrate. The RF signal is provided by a vector network analyser which allows monitoring both the amplitude of the mechanical motion and the phase relation between the driving force and the mechanical response, as shown in Fig. 11.

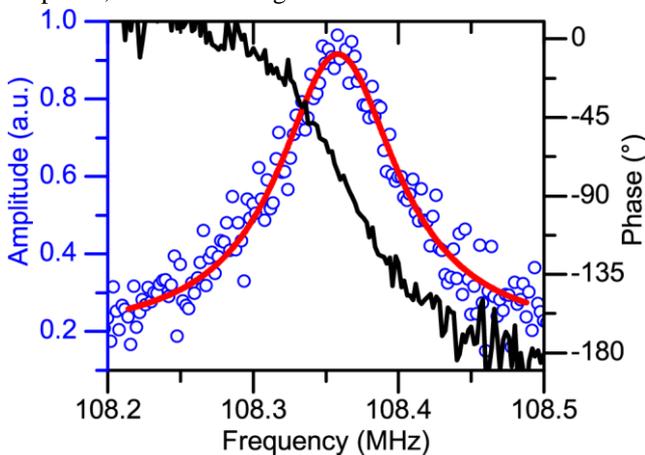

**Figure 11 Electro-optomechanical driven motion:** Amplitude (blue) and phase (black) spectrum for an H-resonator driven at a resonance frequency of 108 MHz. The Lorentzian fit (red) reveals a mechanical quality factor of 1100 in vacuum.

From a Lorentzian fit to the linewidth of the amplitude spectrum the mechanical Q factors of the resonators, excited at their different mechanical resonances, can be extracted. In the frequency range up to 100 MHz quality factors up to 8500 at room temperature and low pressure are found[92], which are slightly reduced when compared to diamond H-resonators without electrical drive.[77] This reduction can be attributed to the mechanical damping caused by the metal electrodes. By employing a different configuration of the electrodes with respect to the mechanical resonator the additional mechanical damping caused by the metal electrode could be prevented.[112]

The electrical drive allows for exciting the motion with higher forces at higher frequencies, which makes it possible to observe the driven motion at ambient pressure. By reducing the device size it becomes possible to operate at frequencies in the GHz regime where air damping is negligible[96,82,88] and even interfacing integrated optomechanics with microfluidics becomes possible.[113] The use of mechanical degrees of freedom enables tunable optical elements and might in the future be interfaced with the sensor concepts discussed in section 5.3.

## 7 Towards single photon optomechanical circuits

Finally we would like to point towards recent directions in nanoscale optomechanics where the properties of diamond may be of benefit, in particular with respect to device operation with single photons. Cavity optomechanical systems are rapidly approaching the regime where the radiation pressure of a single photon is strong enough to cause displacement of the mechanical resonator by more than its zero-point uncertainty. In the framework of optomechanics, the proposals of Bose et al[114,115] provide a route to mechanical states where the effects of unconventional decoherence might be observable. In the single photon strong coupling limit the power spectrum of the cavity output has multiple sidebands which can be discriminated in the resolved sideband limit. When using suitable device parameters, theoretical considerations by Tang and Vitali[116] show that even the force exerted by a single photon on a mechanical resonator may be observed. Current opto-mechanical systems, however, still exhibit couplings below the necessary single-photon interaction strength. As suggested by Ukram et al.[117], this can be alleviated by enhancing the single-photon coupling strength by the presence of a strong pump field as a feasible route towards quantum state transfer between optical photons and micromechanical resonators.

Besides limitations on the optomechanical resonator and the corresponding optical cavity, a suitable single photon generation and readout framework is necessary to observe the system on a monolithic platform. Such a framework will include active elements for generating, manipulating and detecting single photons. Using diamond, single photon generation can be achieved by exploiting color centers. Complementary required are also suitable detectors with sensitivity on the single photon level. One necessary requirement for this measurement is to have detectors with high efficiency and low dark count rates to allow for prolonged integration times, potentially integrated into the optomechanical circuits. Such detectors have been demonstrated in recent years in form of waveguide integrated superconducting nanowire single photon detectors (SNSPDs)[118–121] and could be useful for other suggested implementations of single photon optomechanics.[117,122–124] The geometry of such waveguide integrated SNSPDs is illustrated in Fig. 12. SNSPDs integrated with diamond nanophotonic circuits, fabricated on the same polycrystalline material platform as the optomechanical circuits discussed in section 6, have recently been shown to detect infrared photons at 1550 nm with on-chip detection efficiencies up to 66%[125] when biased close to its critical current, as can be seen in Fig. 12b. These detectors can be operated at an optimal point in terms of lowest noise equivalent power, where the detector still shows an efficiency of 38%, while the dark count rate is below 3Hz.[125]

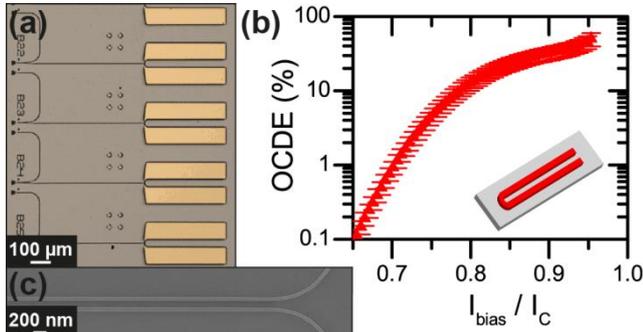

**Figure 12 Single photon detectors**: (a) Optical microscope image of superconducting nanowire single photon detectors integrated with polycrystalline diamond waveguides. (b) On-chip detection efficiency of a SNSPD of 100 nm wire width and 65µm length at varying bias current $I_{bias}$. Maximum detection efficiency of up to 66 % is observed.[125] Inset: layout of the nanowire detector (red) on top of a nanophotonic waveguide (grey). (c) SEM image of a nanowire preform structured into negative resist on top of superconducting niobium nitride and diamond thin films.

While SNSPDs have recently also been implemented on SCD, their efficiency has not been characterized and they have not been integrated into photonic circuits.[126]. In contrast, integrated superconducting nanowire single photon detectors have been fabricated directly on top of PCD waveguides and therefore interface directly with the integrated optical and optomechanical circuits, as described in earlier sections. This allows for combining the rich toolbox of nanophotonics for optical signal processing with using an optomechanical framework for single photon manipulation on chip. In particular, the use of diamond offers the possibility to carry out single photon signal processing over a wide wavelength range, also in the mid- and long infrared spectral regions. Through the combination with free-standing waveguide structures and diamond resonators, such predominantly passive single photon circuits can be equipped with active tunability through optomechanical and electro-optomechanical interactions as pointed out above. The use of the H-resonator in particular allows for implementing compact phase-shifting and beam splitting devices which are essential ingredients for emerging quantum photonic technologies.

**8 Conclusions**

In this article we provided a summary of the research results on monolithically integrating optical and mechanical components on single- and polycrystalline diamond substrates. The use of diamond within the context of optomechanical systems is a natural choice because of outstanding optical and mechanical properties. Recent progress in both regimes has led to the demonstration of both optical and mechanical resonator devices with quality factors in excess of one million as a promising step towards application-driven circuit components. Because diamond can also provide low absorption and very high thermal conductivity, diamond integrated optomechanical devices also hold promise for operation in the high power regime. This may prove advantageous for readout of nanomechanical motion with high precision, as well as for implementing non-linear optomechanical systems on chip. Because diamond offers extremely wide optical transparency currently unexplored operation regimes in the mid and long infrared spectral region become accessible for integrated optomechanical circuits. In particular with respect to sensing and metrology in the important fingerprint spectral region, diamond optomechanical devices might find new applications in mass sensing and chemical analysis.

The unique combination of material properties makes diamond an outstanding system for studying light matter interactions in a wide range of applications. With the advent of mature processing and synthesis routines the necessary diamond thin-film substrates required for optomechanical circuit manufacture have become available. This allows for porting existing nanostructuring recipes available for optomechanical device fabrication to this new material platform. While large area substrates on a waferscale are only available for polycrystalline diamond, recent advances in realizing single crystalline substrates with several millimetres in diameter will make the fabrication of integrated optomechanical devices feasible in the future.[127–129] With advanced transfer and bonding techniques such templates will enable high quality circuit components to be realized with low density of color centers. Using ion implantation for site-specific single photon source realization[130,131] will eventually allow for devising full-fledged diamond optomechanical circuits that operate on a single photon level. In combination with single photon detector technology now available on a diamond platform a flexible framework for the generation, detection and manipulation of single photons will become available in the near future. Such a platform will allow for porting optomechanics at rather high optical intensity to the single photon level with further applications in tunable on-chip quantum photonics.

**Acknowledgements** We acknowledge support by DFG grant PE 1832/1-1 and PE 1832/2-1. P. R. acknowledges support by the Karlsruhe School of Optics and Photonics (KSOP) and the Deutsche Telekom Stiftung. We also appreciate support by the Deutsche Forschungsgemeinschaft (DFG) and the State of Baden-Württemberg through the DFG-Center for Functional Nanostructures (CFN) within subproject A6.4.